\documentclass[conference]{IEEEtran}

\IEEEoverridecommandlockouts
\usepackage{epsfig,color,cite}
\usepackage{amsmath}
\usepackage{algorithm}
\usepackage{algpseudocode}% 
\usepackage{amsthm}
\usepackage{mathrsfs}
\usepackage[center]{caption}
\usepackage{amsfonts}
\usepackage{amssymb}
\usepackage{mathrsfs} 
\usepackage{euscript}
\usepackage{graphicx}
\usepackage{epstopdf}
\usepackage{multirow}
\graphicspath{ {./images/} }

\begin{document}
		\title{Federated Double Deep Q-learning for Joint Delay and Energy Minimization in IoT networks\\
			{\footnotesize \textsuperscript{}}
			\vspace{-11mm}
%			\thanks{ S. Zarandi and H.~Tabassum are with the Lassonde School of Engineering at York University, Canada (e-mail: shz@yorku.ca, hina@eecs.yorku.ca). This work is supported  by the Discovery Grant from the Natural Sciences and Engineering Research Council of Canada.}
		}
		\author{
			\IEEEauthorblockN{Sheyda Zarandi and Hina Tabassum, \textit{Senior Member, IEEE}\thanks{ S. Zarandi and H.~Tabassum are with the Lassonde School of Engineering at York University, ON, Canada (e-mail: shz@yorku.ca, hinat@yorku.ca). This work is supported  by the MITACS research trainee award and Discovery Grant from the Natural Sciences and Engineering Research Council of Canada.} } 
			\vspace{-11mm}
		}
		\raggedbottom
		\maketitle
	\begin{abstract}
	In this paper, we propose a federated deep reinforcement learning  framework to solve  a multi-objective optimization problem, where we consider minimizing the expected long-term task completion delay and energy consumption of IoT devices. This is done by optimizing  offloading decisions,  computation resource allocation, and transmit power allocation.  Since the formulated problem   is a mixed-integer non-linear programming (MINLP), we first cast our problem as a  multi-agent distributed deep reinforcement learning (DRL) problem and  address it using double deep Q-network (DDQN), where the actions are offloading decisions. The immediate cost of each agent is calculated through solving either the transmit power optimization or local computation resource  optimization, based on the selected offloading decisions (actions). Then, to enhance the learning speed of IoT devices (agents), we incorporate federated learning (FDL) at the end of each episode.  FDL enhances the scalability of the proposed DRL framework, creates a context for cooperation between agents, and minimizes their privacy concerns.  Our numerical results demonstrate the efficacy of our proposed federated DDQN framework in terms of learning speed compared to federated deep Q network (DQN) and non-federated DDQN algorithms.  In addition, we investigate the impact  of  batch  size,  network  layers,  DDQN target  network  update frequency on the learning speed of the FDL.
	\end{abstract}
\section{Introduction}
Massive connectivity is among one of the most challenging requirements of Internet-of-Things (IoT) networks which necessitates efficient, scalable, and low-complexity network resource management. Furthermore, due to limited computation and battery capacity of the IoT devices, it is often impossible for them to process their resource-intensive tasks within a predefined deadline. In the sequel, mobile cloud computing (MCC) and  mobile edge computing (MEC) enable IoT devices to offload their tasks to the cloud or edge servers to access their substantial processing capabilities at the expense of having to transmit the tasks over dynamic wireless channels. Subsequently, to take full advantage of the MCC and MEC paradigms, it becomes essential to carefully optimize offloading decisions, communication, and computation resources. 
% Note that, if the task is performed locally, the computation capacity allocated to the task should be carefully determined. On the other hand, if offloading is selected, the transmit power of the devices should be optimized to ensure neither delay nor energy consumption exceeds the acceptable threshold. 

% For example, the amount of energy  an IoT device need to spend on processing a given task can be optimized to improve the performance of device and ensure its tasks' QoS requirement. 

% To date,  wireless resource allocation algorithms leverage on tools from standard optimization which are typically time-consuming  and computationally expensive, thereby not applicable in real-time applications. In the sequel, machine learning (ML) is a potential tool that can help us in making online decisions. 

Most of the existing research works solved  the joint offloading decision, communication, and computation resource allocation problem leveraging on tools from optimization theory \cite{me1,me2}. However, the algorithms were typically non-scalable, time-consuming, and computationally expensive. 
Unlike optimization frameworks, deep reinforcement learning (DRL) enables agents to learn by interacting with the environment. This unique approach to learning, turns DRL into an ideal problem-solving tool in dynamic environments. Yet, most DRL algorithms are centralized  and thus suffer from lack of scalability when the number of devices grow.  Also, the computational complexity of finding an optimal policy may increase exponentially as the state space and action space grow. Furthermore, the centralized learning requires IoT devices to share  their information in order to train the global model which may \textit{violate their privacy} and create \textit{unnecessary communication overhead} on the already scarce frequency spectrum.

% Since each IoT device can process a task locally, on edge server, or on a cloud server,  the complexity of finding optimal offloading decisions would exponentially increase with the number of devices. 
% which are typically time-consuming  and computationally expensive, thereby not applicable in real-time applications. In the sequel, machine learning (ML) is a potential tool that can help us in making online decisions. 

% While many approaches from approximation and reducing the dimension of state and action through unsupervised learning algorithms such as PCA and use of autoencoders have been proposed, such approaches by themselves introduce some level of complexity to the problem solving approach and are not necessarily scalable.

Recently, federated learning (FDL) has emerged as a new paradigm for cooperative learning, where multiple nodes contribute in training a single global model. The devices use their local datasets to train and then offload their local models to the central unit for global aggregation. FDL enhances the cooperation between agents and scalability of the network resource management algorithms. Furthermore, FDL does not require local agents to share their data with any external entity, thereby preserves the privacy of each agent~\cite{fedlearning}.

To date, several research works have explored the problem of minimizing delay and energy consumption of the IoT devices considering a FDL system \cite{1,2,3}. 
\textit{Specifically, in the aforementioned research works,  FDL features were incorporated in the problem formulation and the problem was then solved using traditional optimization theory.} Similarly,  \cite{4,5,6} focused on optimizing different aspects of FDL, such as compression of weights, convergence analysis, reduction in the number of iterations, and incentive mechanisms. In these research works, the capabilities of FDL as a part of \textit{resource allocation solution approach} were not investigated. 
% Instead only in simulations MNIST dataset is utilized to train local models which is not an acceptable representative of dynamics of realistic wireless environment. 
In \cite{7}, the problem of computation resource allocation was addressed considering an FDL system. However, FDL is only considered  to formulate an optimization problem which is later on solved by using a\textit{ centralized} actor-critic agent and without using FDL in the solution approach.

{None of the aforementioned research works applied FDL to enhance the efficacy of solving a realistic wireless resource allocation problem.}  Very recently, \cite{8,9} adopted FDL  to facilitate the learning process in DRL, i.e., local DRL models were trained and then integrated together to cooperatively develop a comprehensive global DRL model. However, in \cite{8}, a cooperative caching scheme was proposed and offloading decisions were not considered. In \cite{9}, computational offloading was considered; however, the network was modeled as a queuing system, transmit power was modeled as an integer variable whose maximum value is equal to the maximum length of the energy queue. 
% Evidently, this approach to transmit power/energy allocation is not precise as these are in fact very fine-grained variables and even subtle changes in their value can considerably change the delay and energy consumption. 
Also, in \cite{9}, no explicit quality-of-service (QoS) was guaranteed for users' tasks and computation resource allocation was overlooked.

	In this paper, we employ federated DRL (FedRL) to address the problem of minimizing joint expected task completion delay and energy consumption of IoT devices with offloading decisions, computation resources, and transmit powers as variables.
	 Considering the mixed-integer non-linear programming programming (MINLP) nature of the problem, we first reformulate our problem as a  multi-agent DRL problem and solve it using  double  deep  Q-network  (DDQN). In this problem, offloading decisions would be the actions and the immediate cost is calculated through solving either the transmit power or local computation resource  optimization. To improve learning quality and speed of DRL, we incorporate FDL  at the end of each {episode}. Using FDL results in a privacy-preserving and scalable framework and creates a context for cooperation between agents.
% These capabilities are specially important in IoT networks, where on one hand the large number of devices necessitates a distributed and scalable resource allocation scheme, and on the other hand, due to the typically limited resources on devices, extensive learning tasks cannot be performed locally. 
Our numerical results demonstrate the efficacy of our federated DDQN framework in terms of learning speed compared to federated deep Q-network (DQN) and non-federated DDQN algorithms.

 The rest of this paper is organized as follows: Section II describes the system model and the problem is formulated in Section~III. Then, in Section IV, our proposed algorithm is presented. Finally, simulation results and related discussions are provided in Section V followed by conclusions in Section~VI.

\section{System model and Assumptions}
	We consider a network containing one MEC server, one cloud server, and a set of $\mathcal{N}=\{1,...,N\}$ IoT devices with limited computation and energy resources. {We consider a given time horizon $\mathcal{T}$ which is divided into $T$ time steps.} At each time $t$, device $i$ needs to process one of the tasks in its queue, defined with the tuple $(L_{i,t},C_{i,t},\hat{T}_{i,t})$, where $L_{i,t}$ is the size of the task (in bits), $C_{i,t}$ denotes the CPU cycle requirement of the task, and $\hat{T}_{i,t}$ denotes the maximum delay threshold of the task. At any time  $t$, devices can either execute their task locally or offload it to edge or cloud server. 
	
	Let us denote local offloading decision of device $i$ at time $t$ as $x_{i,t}$, where $x_{i,t}=1$ means the task would be performed locally, and $x_{i,t}=0$, otherwise. Similarly, we define MCC and MEC offloading variable of device $i$ by $z_{i,t}$ and $y_{i,t}$, respectively. If device $i$ offloads its task  to the cloud $z_{i,t}=1$ and if it offloads the task to the edge server $y_{i,t}=1$. As a binary offloading decision is considered, we have:
	\begin{equation}
		x_{i,t}+y_{i,t}+z_{i,t}=1.~\forall{t}\in \mathcal{T}.
	\end{equation}  
	When device $i$ offloads its task (whether to MEC server  or to the cloud), the delay and energy consumption would depend on the channel condition, the size of the task, and the power with which the device transmits its task and, in the case of local computation they depend on computation resource utilization. In what follows, we model the delay and energy consumption that an IoT device would experience, given its offloading decision. 
	
	If the device $i$ decides to offload its task, it should first transmit it to the MEC-enabled base station (BS) through wireless channels. At time step $t$, the transmission data rate of this user, denoted by $r_{i,t}$ is calculated as follows:
	\begin{equation}
		r_{i,t}=B\log_2(1+\frac{p_{i,t} h_{i,t}}{\sigma^2}),
	\end{equation}
	where $B$ and $p_{i,t}$ denote the bandwidth and transmit power of device $i$ at time step $t$, respectively. Also, $h_{i,t}$ and $\sigma^2$ represent the path-gain of device $i$ at time $t$ and the receiver noise.
	Thus, the communication delay and energy consumption of device $i$, while offloading is given, respectively, as follows:
	\begin{equation}
T_{i,t}^{\mathrm{comm}}=\frac{L_{i,t}}{r_{i,t}},
	\end{equation}
	\vspace{-4mm}
	\begin{equation}
E_{i,t}^{\mathrm{comm}}=p_{i,t} T_{i,t}^{\mathrm{comm}}=\frac{p_{i,t} L_{i,t}}{B\log_2(1+\frac{p_{i,t} h_{i,t}}{\sigma^2})}
	\end{equation} 
	If device $i$ offloads its task to the edge server the computation delay would be
$
T^e_{i,t}=\frac{C_{i,t}}{F^e},
$
where $F^e$ denotes the average computation capacity of edge server.
Also, if device $i$ offloads the task to cloud server, the computation delay would be
$
T^c_{i,t}=\frac{C_{i,t}}{F^c},
$ where $F^c$ represents the average computation capacity of the cloud server.

From the perspective of IoT device, the energy that is consumed for processing a task when it is offloaded to either of the servers, is the energy spent on the transfer of the task. Therefore, both cloud and edge computing energy utilization at step $t$, denoted by $E^c_{i,t}$ and $E^e_{i,t}$, would be equal to 
$E_{i,t}^{\mathrm{comm}}$.

% 	\subsubsection{Delay and Energy Consumption Model (No Offloading)}
	If device $i$ chooses to perform its task locally, the local computation delay and energy consumption would depend on the amount of computation resource  allocated to process the task at time $t$, which we denote by $f_{i,t}^L$.  Thus, the local delay and energy consumption of the device is modeled as follows:
	\begin{equation}
T_{i,t}^{L}=\frac{C_{i,t}}{f_{i,t}^L}, ~~ E_{i,t}^L= \kappa_i(f_{i,t}^L)^2,
	\end{equation}

where $\kappa$ is a constant coefficient that depends on the chip architecture in devices. Note that higher resource utilization (transmit power or computation capacity) , decreases the task completion delay at the expense of increased energy consumption. Therefore, this trade-off must be carefully managed through efficient offloading decision making and precise optimization of {$f_{i,t}$ and $p_{i,t}$ } in the case of local computation and offloading, respectively.

\section{Multi-objective Problem Statement}
In this section, we formulate the multi-objective problem of jointly minimizing the long-term delay and energy consumption of an IoT device in a decentralized manner over a specified time horizon $\mathcal{T}$. The long-term expected cost (weighted sum of delay and energy consumption) for each IoT device $i$ is formulated, respectively, as follows:
\begin{align}\label{delay}
T_{i}(\mathbf{p}_i,&\mathbf{f}_i,\mathbf{x}_i,\mathbf{y}_i,\mathbf{z}_i)= \mathbb{E}[\lim_{t\rightarrow \infty} \frac{1}{t}\sum_{j=0}^t  \left(x_{i,j} T_{i,j}^L+\right.\nonumber \\ & \left. y_{i,j}(T^e_{i,j}+T_{i,j}^{\mathrm{comm}})+ 
z_{i,j} (T_{i,j}^c+T_{i,j}^{\mathrm{comm}}+\Psi)\right)],
\end{align}
\small
\begin{align}\label{energy}
&E_i(\mathbf{p}_i,\mathbf{f}_i,\mathbf{x}_i,\mathbf{y}_i,\mathbf{z}_i)\nonumber\\&=\mathbb{E}[\lim_{t\rightarrow \infty} \frac{1}{t} \sum_{j=0}^t \left(x_{i,j} E_{i,j}^L+ y_{i,j} E^e_{i,j} +z_{i,j} E^c_{i,j}\right)],
\end{align}
\normalsize
where $\mathbf{p}_i=[p_{i,1},p_{i,2},\cdots,p_{i,t}]$, $\mathbf{f}_i=[f_{i,1},f_{i,2},\cdots,f_{i,t}]$, $\mathbf{x}_i=[x_{i,1},x_{i,2},\cdots,x_{i,t}]$, $\mathbf{y}_i=[y_{i,1},y_{i,2},\cdots,y_{i,t}]$, and $\mathbf{z}_i=[z_{i,1},z_{i,2},\cdots,z_{i,t}]$, represent the vectors of transmit powers, computation resource allocation, local computing, edge offloading, and cloud offloading decision of device $i$, respectively. As cloud server is generally located far from the IoT devices, the delay of accessing cloud is commonly more than the delay of offloading to the edge server which is located at the edge of the network. {In the equation (\ref{delay}),  $\Psi$ denotes the delay of accessing the cloud server, including the time necessary to transfer the task from BS to the cloud, the possible routing in the path, and the response delay.}

For any given device $i$ at time $t$, we model our problem as:
\begin{equation}\label{first}
\begin{aligned}
&\min_{\mathbf{p}_i,\mathbf{f}_i,\mathbf{x}_i,\mathbf{y}_i,\mathbf{z}_i}~ T_i+\lambda_i E_i \\
&~~\text{subject to:}\\
& \mathrm{C1}: f_{i,t}^L<= F_{\mathrm{max},i},~~ \forall t \in \mathcal{T}, \\
& \mathrm{C2}:x_{i,t} E^L_{i,t}+y_{i,t} E^e_{i,t}+z_{i,t} E^c_{i,t}<= E_{\mathrm{max},i},~~\forall t \in \mathcal{T},\\
& \mathrm{C3}: T_{i,t}<= \hat{T}_{i,t}, ~\forall t \in \mathcal{T},\\
& \mathrm{C4}: x_{i,t}+ y_{i,t}+z_{i,t}=1,~~\forall i \in \mathcal{N}, \forall t \in \mathcal{T},\\
& \mathrm{C5}: x_{i,t},y_{i,t},z_{i,t} \in \{0,1\},~~\forall i \in \mathcal{N},\forall t \in \mathcal{T}.\\
\end{aligned}
\end{equation}

In the above optimization problem, $\lambda_i$ is a weighting factor whose value should be carefully selected based on the heterogeneity of resources available at each individual IoT device. If device $i$ is more restricted in the energy resource compared to computation resource, the value of $\lambda_i$ should be set to a larger number. Otherwise, $\lambda_i$ should be  a small number. Furthermore, constraint C1 indicates the local computation capacity with  the maximum threshold $F_{\mathrm{max},i}$. Constraint C2 represents the restriction on the energy resource of the device and that energy utilization should not exceed $E_{\mathrm{max},i}$. Furthermore, constraints C4 and C5 define the binary offloading scheme adopted in this paper. It can be proven that both equations (\ref{delay}) and (\ref{energy}) are convex with respect to the variables $p_{i,t}$ and $f_{i,t}$, respectively. However, with binary offloading variables ($x_i$, $y_i$, and $z_i$) included, (\ref{first}) turns into a MINLP that cannot be solved in an acceptable time span.

\section{Proposed Federated DDQN Algorithm}
To solve (\ref{first}) at each IoT device, we propose a DDQN algorithm, and solve the problem in the following two phases: 

$\bullet$ \textit{Offloading Decision Optimization:} 
	Since each IoT device has three options to process a task (namely local, edge server, or cloud server computing), there are almost $2^{3N}$ possible offloading options (from the perspective of a centralized controller) at each given time step. As the number of devices increases, this complexity would also surge exponentially. To address this problem,  we apply a multi-agent DDQN framework where each IoT device would train their local DDQN models using their local data. 
	
$\bullet$ \textit{Computing and Communication Resource allocation:} Given the offloading decision, we optimize computation capacity or transmit power of the devices to minimize the weighted sum of energy consumption and delay. 
	We use optimization theory to address this part of the problem and then feed the results into the DDQN framework as the immediate cost function. In this way, we provide the learning agent with a real sense of the quality of the adopted offloading policy that reflects many important aspects of the system model (such as limitation of resources in each device and their QoS demands).

After the DDQN agent is trained through the above mentioned process for one training round, we apply a federated learning framework where each IoT device will train its DDQN models, share their models with the centralized controller, and update their models to the central aggregating unit. This mechanism is detailed  in the flowchart provided in Fig.~1.

In what follows, we first focus on developing local models through DDQN algorithm and then explain how FDL would be deployed.

% to share the gained knowledge of each agent with others, avoid the need for constant transfer of large amount of data which endangers both privacy and spectral efficiency, and to enhance scalability of the approach even further, we perform FDL. 

\subsection{Double deep Q-network for offloading decision making}
In the first step of our algorithm, we model our problem as a multi-agent DDQN problem. For each device (DRL agent), we have following components:
\begin{enumerate}
    \item \textit{State space}: the state space for each agent $i$, denoted by $s_{i}$, consists of the following components: the length of the  task queue of device $i$ (tasks that are not yet processed or are not successfully processed, would be kept in this queue) which is denoted by $\mathcal{L}_{i,t}$, the path gain of the IoT device $h_{i,t}$, the size of the task currently being processed $L_{i,t}$, its CPU cycle requirement, $C_{i,t}$, and available resources. Thus,
        $s_{i}=\{\mathcal{L}_{i,t},h_{i,t},L_{i,t},C_{i,t}, E_{\mathrm{max},i},F_{\mathrm{max},i}\}$.
    If a task is successfully processed {under a given offloading decision policy} (its  QoS requirement is satisfied), it would be removed from the task queue of the device. Otherwise, it will remain at the top of the queue to be processed {under another offloading decision}.
    \item \textit{Action space}: The action space of agents, denoted by $\mathcal{A}$, contains possible offloading decisions, i.e.,  whether to process the task locally or offload.
    \item \textit{Cost}:  (\ref{first}) suggests that the cost of an agent  is equal to the weighted summation of the delay and energy consumption given in the objective function. The value of this objective function and thus the cost  depends on the value of $p_{i,t}$ in the case of offloading and $f_{i,t}$ if local computation is selected. Therefore, to ensure that the cost function accurately reflects the benefit of a given offloading decision, these variables should be carefully optimized. To this end, when local computation is selected ($x_{i,t} = 1$), the cost would be calculated by solving the instantaneous optimization problem below:
    \begin{equation}\label{comp_opt}
\begin{aligned}
&~\min_{{f}_{i,t}}~ T_{i,t}^L+\lambda E_{i,t}^L \\
&\text{subject to: C1, C2, C3.}
\end{aligned}
\end{equation}
in the case offloading is selected ($y_{i,t}=1$ or $z_{i,t}=1$), the transmit power would be optimized by solving the following optimization problem:
\begin{equation}\label{power_opt}
\begin{aligned}
&~\min_{{p}_{i,t}}~ y_{i,t}(T_{i,t}^e+T_{i,t}^{\mathrm{comm}}+\lambda E_{i,t}^{\mathrm{comm}})\\&~~~~~~~~~~~~~+ z_{i,t}(T_{i,t}^c+T_{i,t}^{\mathrm{comm}}+\Psi+\lambda E_{i,t}^{\mathrm{comm}}) \\
&\text{subject to: C2, C3.}
\end{aligned}
\end{equation}
\end{enumerate}

The design of state and cost function has a significant impact on the success of DDQN in finding the optimal {offloading} policy, $\pi^*$. 
By using a multi-agent approach, we are in fact limiting the state and action space and focus on each device separately. Also, by modeling the cost function as an optimization problem not only can we optimize the local resource utilization and enforce system constraints, but also we can provide the agent with an accurate estimation of the quality of an offloading decision. Note that it can be easily proved that both (\ref{comp_opt}) and (\ref{power_opt}) are convex single variable optimization problems that can be solved using standard  softwares.
% Therefore, on one hand solving them is not time consuming or computationally exhaustive and on the other hand can incorporate many aspects of the environment and derive an accurate estimation of state-action value.

% All devices aim to learn the optimal policy that maximizes their cost. Meaning at each state $s$, device $i$ would choose action $a_i$ that maximizes its returns.
\begin{algorithm}
\caption{ Proposed Federated DDQN Algorithm}
\label{euclid}
\begin{algorithmic}[1]
	\State~~Initialize the global model $\theta^{\mathrm{global}}$, and set maximum FDL iterations to $K$.
	\State~~For each agent initialize online and target networks as: $\theta_{i}^{\mathrm{online}} = \theta_{i}^{\mathrm{target}} =  \theta^{\mathrm{global}}$.
	\State\textbf{While} ($K \geq 0$):
	\State~~~Set $\theta_{i}^{\mathrm{online}} = \theta_{i}^{\mathrm{target}} =  \theta^{\mathrm{global}}$,  $\forall i \in \mathcal{N}$,
	\State~~ Select the set of participating devices, $\mathcal{I}$, based on (\ref{device_selection}),
	\State~~ \textbf{For} each device $i$ in $\mathcal{I}$ \textbf{do}:
	\State~~~~~~ \textbf{For} each time step $t$ if $|\mathcal{L}_{i,t}| > 0$ \textbf{do}:
	\State~~~~~~~~ Interact with environment and calculate the cost using (\ref{comp_opt}) or (\ref{power_opt}),
	\State~~~~~~~~ Save the experience in replay memory $\mathcal{M}_{i,t}$.
	\State~~~~~~~~ Train the local model on $\mathcal{M}_{i,t}$, 
	\State~~~~~~~~ Transmit $\theta_{i}^{\mathrm{online}}$ to the central aggregation unit,
	\State~~~~~~ \textbf{End For}
	\State~~ \textbf{End For}
	\State~~ update $\theta^{\mathrm{global}}$ using (\ref{aggregate}).
\end{algorithmic}
\end{algorithm}

Let us denote the immediate cost of each device $i$ obtained from the solution of the above mentioned optimization process as  $u_i(s,a)$. Using Bellman equation, the action-state value is:
\begin{equation}
    Q_i({s,a})=u_i(s,a)+\gamma \sum_{s'\in \mathcal{S}} P_{ss'}(a) \max_{a'\in \mathcal{A}} Q_i(s',a'),
\end{equation}
where $\mathcal{S}$, $P_{ss'}(a)$, and $\gamma$ are the set of states, the transition probability function, and the discount factor, respectively. To overcome the need for having a full model of environment, calculating the transition probability function, and to acquiring a more stable learning process, DDQN is employed in this work. Each agent $i$ has two neural networks working alongside each other, one called \textit{online network} with parameters $\theta_i^{\mathrm{online}}$ and the other called \textit{target network} with parameters $\theta_i^{\mathrm{target}}$. At each training iteration the target value for training the online network in device $i$ is calculated as:
\begin{equation}
    L_i= u_i(s,a)+\gamma Q_i(s',\arg \max_{a'\in \mathcal{A}} Q_i(s',a';\theta_i^{\mathrm{online}}),\theta_i^{\mathrm{target}})
\end{equation}
While $\theta_i^{\mathrm{online}}$ is updated at every iteration, the frequency of change in $\theta_i^{\mathrm{target}}$ is typically much lower and only once in every $f_\mathrm{update}$ rounds, $\theta_i^{\mathrm{target}}$ would be set equal to $\theta_i^{\mathrm{online}}$.

% \subsection{Motivation of Federated Reinforcement Learning}
% One of the major disadvantages of DRL, especially in IoT networks in which various devices with heterogeneous computation and energy capacity exist, is the fact that training a DRL agent can be quite computationally exhaustive. As such it is often infeasible to train a DRL model on resource-constrained IoT devices.

% As mentioned, using a central agent to gather information from network devices and train a global model is also problematic as 1) this approach is not scalable as a global view of the system requires considering the information of all devices and their actions which make the approach impractical to use in dense IoT networks, 2) some components of state space may be privacy sensitive (for example location), in which case devices may not willingly share and disclose them, and 3) a great amount of data should be constantly transferred between devices and the central unit, which results in over utilization of already scarce network resources such as bandwidth.

As discussed before, training a DRL agent in a centralized manner can lead to critical issues related to scalability, agents' privacy, and additional communication overheads. On the other hand, training a DRL agent in a distributed  manner can impact the overall performance gains (e.g., an agent might consume a longer time to train its model).
As such, we consider FDL to combine the benefits of both centralized and distributed learning. FDL enables each agent to train its own local model, using its own local data. Then these local models are sent to a central aggregation unit to be combined together. This process continues until a criterion is met.
% Combining DRL and FDL together results in numerous advantages for IoT networks. As IoT devices have limited resources, the cooperative learning scheme available in the aggregation step of FDL relieves these devices from having to develop their own models from scratch and all by using their own limited resources and data. Furthermore, as the DRL agent does not have to adopt a global view, the need to consider all IoT devices, their states and actions, would be obliterated, thereby reducing the size of the action space and facilitating the training process. Also, since agents do not need to share their experiences with any external entity, not only the privacy of agents would preserved, but also the need for frequent transfer of high volume information would be rectified.  This motivates us to employ federated DDQN approach to address our problem.

\begin{figure}[h]
    \centering
    \includegraphics[width=80mm,height=90mm, angle=270]{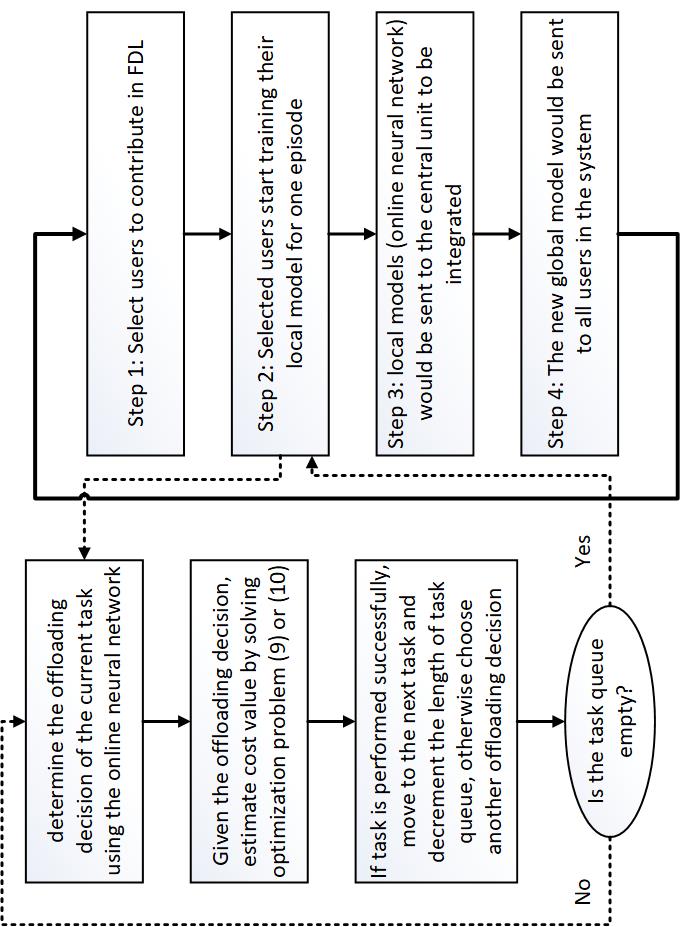}
    \caption{The federated reinforcement learning process}
    \label{flowchart}
\end{figure}
\vspace{-5mm}
\subsection{Federated DRL Approach}
The steps to train the FedRL agents are presented in the following:
\subsubsection{Device selection strategy}
At the beginning of each iteration of FDL, a set of IoT agents are selected to participate in the FDL. Thus, of all $N$ devices in the network, only a small subset, denoted by $\mathcal{I}=\{1,...,I\}$ is selected to contribute in FDL. In this paper, the device selection is done based on the following criterion:
\begin{equation}\label{device_selection}
     \arg \max_{i\in \mathcal{N}}~\text{Var} ( \frac{d_{i}P_{\mathrm{max},i}}{F_{\mathrm{max},i}}),
\end{equation}
where $d_i$ represents the distance of device from BS and the function Var stands for variance. This criterion helps in identifying devices whose experiences are more heterogeneous and thus can  contribute more in the the learning process.

\subsubsection{Training local models}
As explained previously , all IoT devices use DDQN to train their local models. After this local training is finished (no more unprocessed task remains in the queue), the weights of online network, $\theta_{i}^{\mathrm{online}}$, is extracted in each agent and is then sent to the central aggregating unit. 

\subsubsection{Model Aggregation}
When central unit receives the models of participating IoT devices, it would aggregate the models which results in a single global model that would be then transmitted to all agents. For the purpose of aggregation, we utilize FedAvg \cite{fedavg}, and perform model aggregation as:
\begin{equation}\label{aggregate}
    \theta^{\mathrm{global}} = \frac{\sum_{i\in \mathcal{I}}\theta_{i}^{\mathrm{online}}}{|\mathcal{I}|}.
\end{equation}

This global model, which has integrated the experiences of all devices, is then transmitted back to IoT devices and the three steps above would be repeated. The details of our proposed framework is provided in \textbf{Algorithm 1} as well as the flowchart given in Fig. 1.

\vspace{-3mm}
\section{Simulation results}
Here, we present our simulation results and extract useful insights related to the performance of our proposed federated DDQN framework in comparison to federated DDQN and distributed DDQN algorithms. In addition, we investigate the impact of batch size, network layers, target network update frequency on the convergence of the FDL. In what follows, we first focus on the impact of parameters of DRL on the learning speed of our proposed algorithm and then the comparison of the proposed algorithm with benchmarks would be presented.

To simulate our system, we consider a network of 100 IoT devices among {which only 20 devices are selected in each round} to contribute in the FDL process. Without loss of generality and for the sake of fair comparison, we assume the maximum computation capacity and energy consumption limit of the IoT devices are  1 Gbps and  23 dBm, respectively. 

Fig.~2 demonstrates the effect of network architecture on the convergence of our proposed FedRL algorithm. Here we have some shallow networks with up to five layers and deeper networks that are obtained by stacking multiple layers with [16,32,32] neurons on each other. We note that by increasing the number of layers, faster model training can be achieved. The reason behind this observation is that, by exploiting deeper neural networks, we can better find the patterns in data (here devices' experiences), which subsequently improves the quality of our local models. Thus, the global model is trained much faster  as its underlying components, local models, are more accurate.

However, since our algorithm will be executed on IoT devices that often lack necessary resources to train a deep network, it may be infeasible to implement deeper neural networks. Therefore, in the next figure, we select a rather simple network architecture with [30,64,16,32,32] neurons in each layer and instead look for other parameters that may facilitate the learning process.
\begin{figure}
    \centering
    \label{fig:arch}
            \vspace{-5mm}
\includegraphics[width=\linewidth, height=6cm]{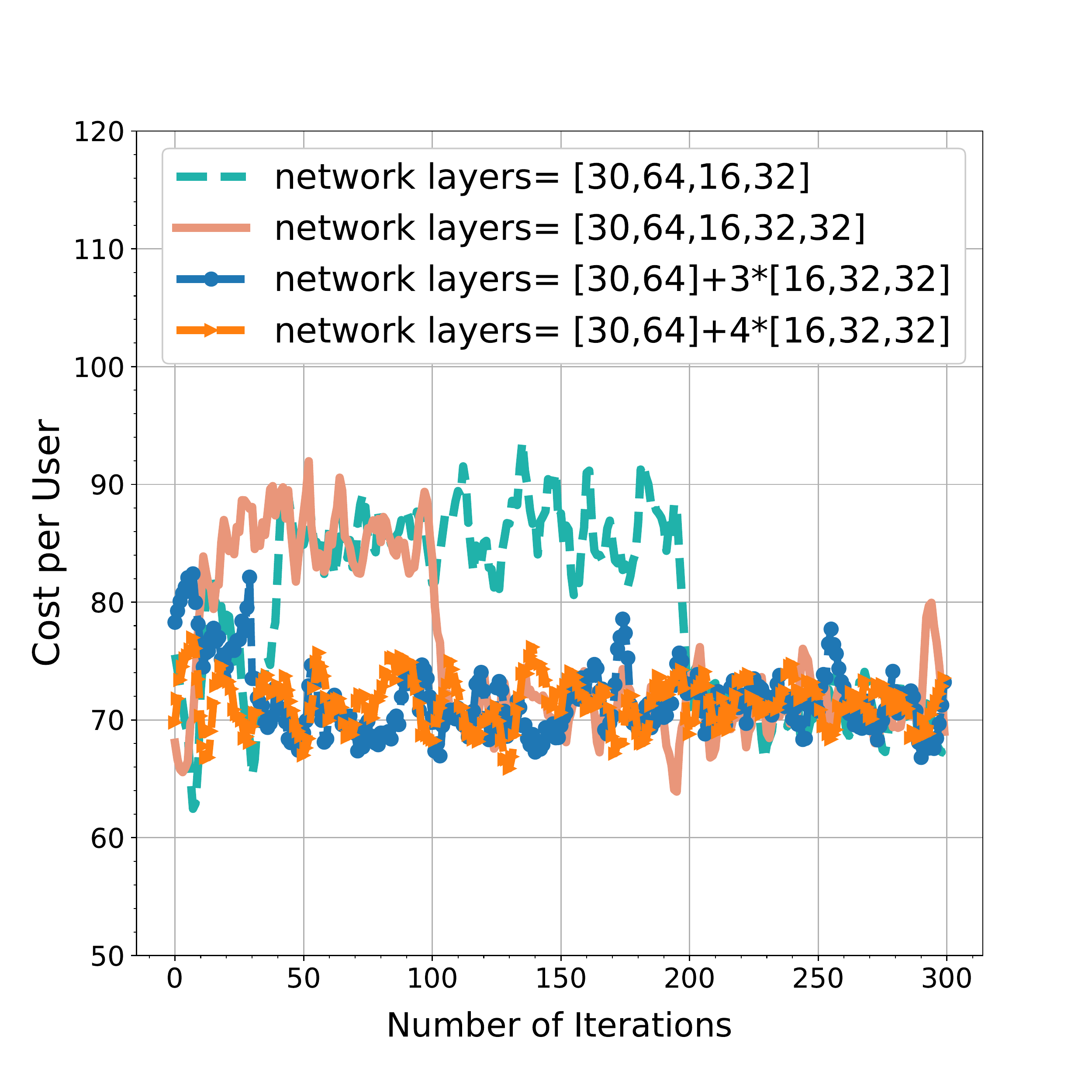}
\caption{The impact of neural network architecture on the convergence of the proposed FedRL algorithm.}
\vspace{-5mm}
\end{figure}

The other parameter we focus on in Fig. 3, is the batch size. We can observe from this figure that as the batch size increases the convergence of our proposed FedRL algorithm becomes faster. When batch size is equal to 10 and is extremely small it takes up to 200 iterations to finally converge to a relative global model, whereas in case batch size is 30, convergence is achieved almost around iteration number 40. By increasing the size of batches, we are basically training our model using more data instances. which results in enhancing the quality of local models and faster training process. 
\begin{figure}
    \centering
    \label{fig:batch}
\includegraphics[width=\linewidth, height=6cm]{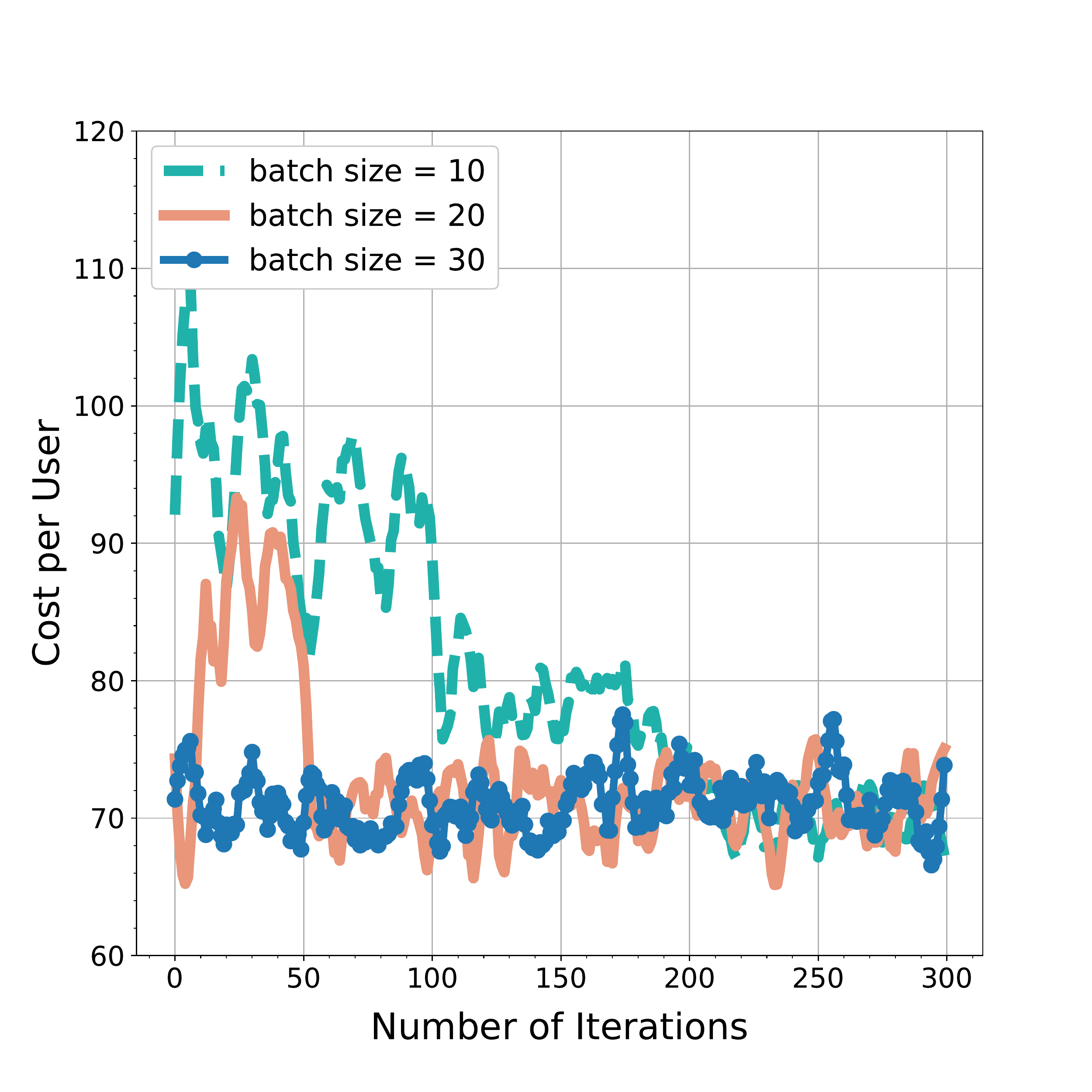}
\caption{The impact of batch size on the convergence of the proposed FedRL algorithm.}
\vspace{-7mm}
\end{figure}

Similar to network architecture and the stated concern regarding the limited computation capacity, memory is another bottleneck in learning process of IoT devices. Larger batch size means higher memory consumption. \textit{If the device is limited both in CPU and memory capacity, neither very deep neural networks nor increased batch size can be a proper solution to facilitate deployment of FedRL on IoT devices.}

To this end, in Fig. 4, we illustrate the effect of one of the parameters of DDQN, namely frequency of updating target network with online network. We can observe here that while the effect of this parameter on performance of DDQN algorithm is well investigated, this parameter is also considerably effective in the performance of federated DDQNs. Since many of the components in our state space, such as path-gain, QoS of tasks, and the length of the task queue, are constantly changing, efficient choice of the frequency of updating target network can stabilize the environment enough for the agent to track it better and obtain a better solution. This effect on local models is quite notable on the FDL as well.
\begin{figure}[t]
    \centering
    \label{fig:freq}
\includegraphics[width=\linewidth,height=6cm]{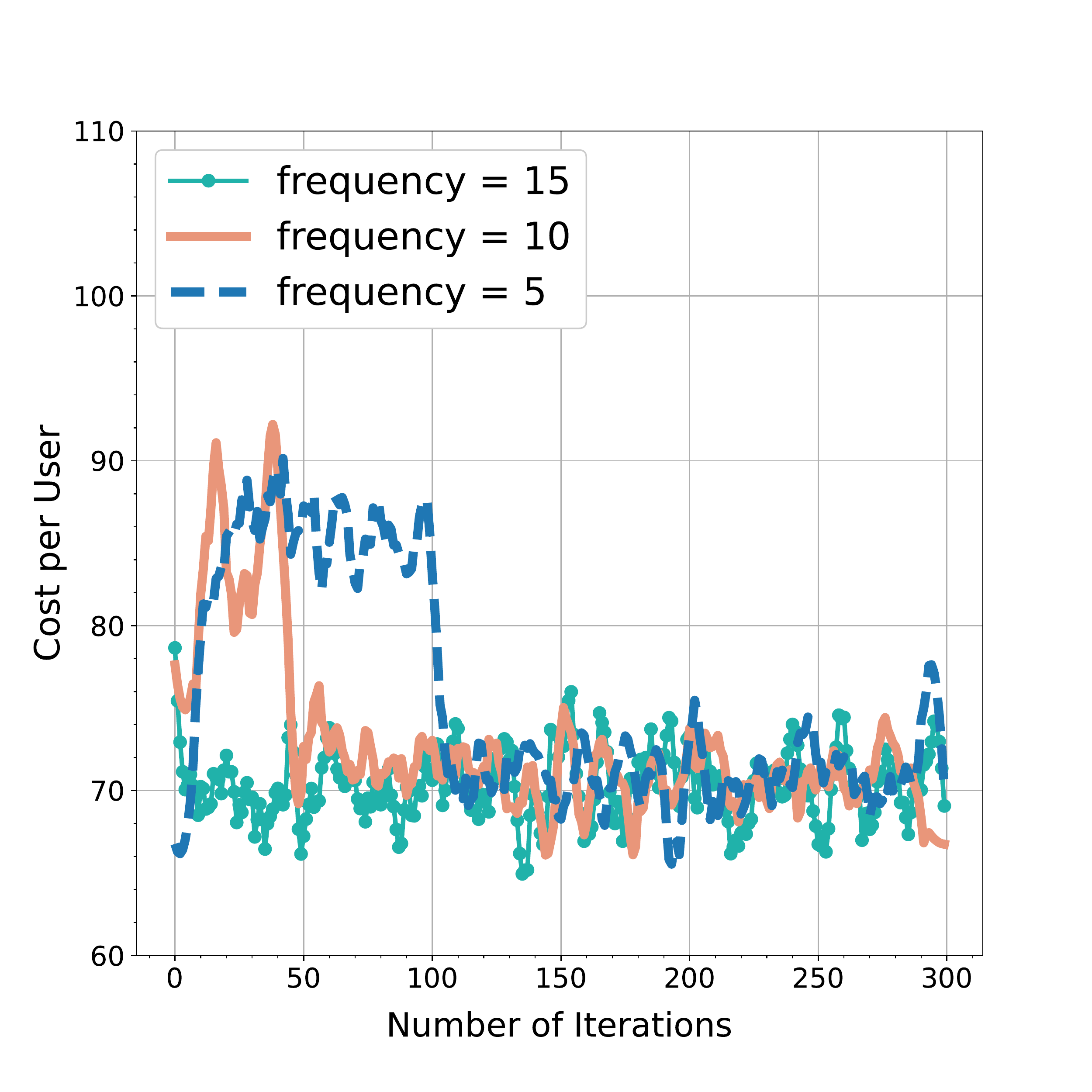}
\caption{The impact of target network update frequency on the convergence of the proposed FedRL algorithm.}
\vspace{-5mm}
\end{figure}

In Fig. 5, we compare the performance of our proposed federated DDQN approach with those of federated DQN and simple distributed DDQN without any aggregation. As can be seen, the performance of federated DDQN is superior to federated DQN in terms of learning speed. As previously explained, the main advantage of DDQN over DQN is the capability to keep target network stationary, helping with the tractability of states' values and subsequently a faster convergence to the correct estimation of them. The impact of this approach is even more notable when DRL is combined with FDL, since if the local models are not correctly trained, their errors would be propagated to other devices' local model through aggregation. Therefore, aggregation can in fact negatively effect the result. 
% Clearly,  the performance of simple distributed learning that does not use any aggregation between local models is on average much better than federated DQN. It is only after many iterations that the performance of these two algorithms get close to each other.

The comparison between distributed DDQN and federated DDQN underlines that the benefits of federated DRL are not limited to its scalability and privacy preservation. The aggregation incorporated in FDL provides IoT devices a great context to cooperatively train their models and merge their intelligence together while preserving privacy of their information. Exploiting federated learning, at every training round  is almost as if devices' models are trained with $I$ times more data than their local information. The significance of this share of knowledge and not data is quite notable in Fig. 5, where as the result of this aggregation step, federated DDQN is working much better and faster than simple distributed DDQN. Note that, averaging is used as the smoothing function.
\begin{figure}
    \centering
    \label{fig:my_label}
\includegraphics[width=8cm, height=6cm]{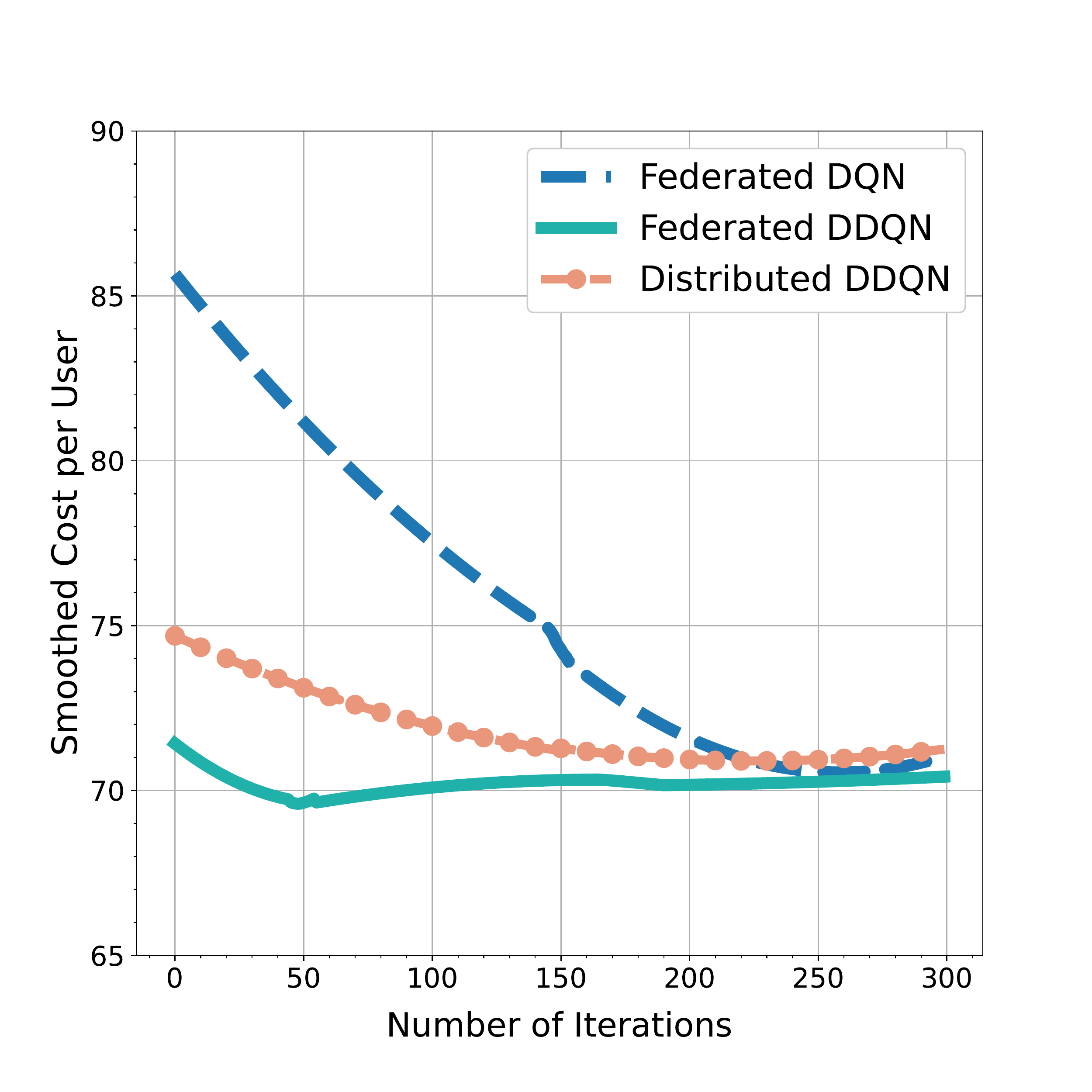}
\caption{{Performance of federated DDQN compared to federated DQN and distributed non-federated DDQN.}}
\vspace{-3mm}
\end{figure}
\vspace{-2mm}
\section{Conclusion}
In this paper, we investigated the problem of joint delay and energy minimization in an IoT network with a three-tier offloading scheme. To solve this problem we combined FDL, DDQN, and optimization theory. Combination of these tools helped us to achieve a scalable, privacy-preserving, and computationally efficient framework for joint power and computation resource allocation and offloading decision optimization. In simulation results, we compared our work with those of 1) federated DQN to demonstrate the superiority of DDQN, especially in dynamic environments and 2) with distributed DDQN to signify the impact of aggregation step incorporated in FDL on the performance of the framework.
\vspace{-2mm}
\bibliographystyle{IEEEtran}
\bibliography{ref}
\end{document}